# Possible evidence for electromagnons in multiferroic manganites

A. Pimenov[1*], A.A. Mukhin[1,2], V.Yu. Ivanov[2], V.D. Travkin[2], A.M. Balbashov[3], and A. Loidl[1]

[1] Experimentalphysik V, Center for Electronic Correlations and Magnetism, Universität Augsburg, Universitätsstr. 2, 86135 Augsburg, Germany
[2] A. M. Prokhorov General Physics Institute of the Russian Acad. of Sciences, 119991 Moscow, Russia
[3] Moscow Power Engineering Institute, 105835 Moscow, Russia
* e-mail: andrei.pimenov@physik.uni-augsburg.de

**Magnetodielectric materials are characterized by a strong coupling of magnetic and dielectric properties and in rare cases simultaneously exhibit both, magnetic and polar order. Among other multiferroics, $TbMnO_3$ and $GdMnO_3$ reveal a strong magneto-dielectric (ME) coupling and as a consequence fundamentally new spin excitations exist: Electro-active magnons, or electromagnons, i. e. spin waves which can be excited by ac electric fields. Here we show that these excitations appear in the phase with an incommensurate (IC) magnetic structure of the manganese spins. In external magnetic fields this IC structure can be suppressed and the electromagnons are wiped out, thereby inducing considerable changes in the index of refraction from dc up to THz frequencies. Hence, besides adding a new creature to the zoo of fundamental excitations, the refraction index can be tuned by moderate magnetic fields, which allows the design of a new generation of optical switches and optoelectronic devices.**

Recently enormous progress has been made in the field of multiferroics and the discovery of new classes of ferroelectro-magnets (FEMs) with the simultaneous occurrence of magnetic and polar order [1,2,3,4,5] has triggered a revival [6,7] of this old field of magneto-dielectric effects and the electrodynamics of multiferroic media [8]. Besides promising applications of FEMs in the field of modern electronics, e. g. as multiple-state memory devices with mutual magnetic or electric control or as magnetically switchable optical devices, fascinating new problems can be tackled in basic research, like the search for magneto-dielectric excitations. The existence of elementary excitations due to the ME interaction, at that time termed Seignette-Magnons, has been theoretically predicted 35 years ago [9]. Here we report the first possible observation of new hybrid excitations in $GdMnO_3$ and $TbMnO_3$, namely magnons which can be excited by an ac electrical field and can be used to fine-tune the index of refraction by moderate magnetic fields.

To demonstrate these effects we choose the two multiferroic manganites, $GdMnO_3$ and $TbMnO_3$. Systematic investigations of the magnetic [10,11] and multiferroic properties [1,2,12,13] of the rare earth manganites $RMnO_3$ ( R = Gd, Tb, Dy) revealed a transition from a paramagnet (PM) into an incommensurate antiferromagnet (IC-AFM) and subsequently into a canted-antiferromagnetic (CA-AFM) structure in $GdMnO_3$ or a commensurate antiferromagnetic (C-AFM) phase in $TbMnO_3$ and $DyMnO_3$. At the lock-in transition from the IC- to the C-AFM, ferroelectricity is induced [1,2,12]. Recent neutron scattering experiments revealed [14] that the lock-in transition in $TbMnO_3$ rather corresponds to a transition into a non-collinear incommensurate magnetic structure.

Figure 1 represents magnetic-field induced changes in the THz-dielectric properties of $GdMnO_3$ and $TbMnO_3$ at selected temperatures close to and below the IC-CA transition. The insets in both frames provide schematic (B,T) phase diagrams for B||c [11,12]. One significant difference is that the CA-AFM phase in $TbMnO_3$ is shifted to much higher fields and the low-temperature phase is a modulated magnetic and ferroelectric [12,14]. In $GdMnO_3$ the canted AFM extends almost to zero fields and a metastable phase exists for 9K < T < 18K. At the magnetic field-induced transition from the modulated to the CA-AFM structure both, the dielectric constant and the dielectric loss reveal significant step-like changes in both compounds. In $GdMnO_3$ the transition reveals almost no hysteresis at temperatures above 30 K, but hysteresis effects increase approaching temperatures close to 18 K. For 9 K < T < 18 K the incommensurate antiferromagnetic state is not recovered after removing the external magnetic field and the sample remains a canted antiferromagnet. Therefore, in this range of temperatures both the canted AFM and the modulated phase are energetically very close and obviously are in metastable equilibrium. For T = 15 K, ε*(H) is typical for this metastable range and can be obtained within a zero-field-cooling (ZFC) cycle. After switching to the CA-AFM state the sample remains in this state independent upon the subsequent magnetic-field history. The application of a negative magnetic field can only reverse the direction of the effective magnetic moment, but *no changes* are observed in dielectric properties.

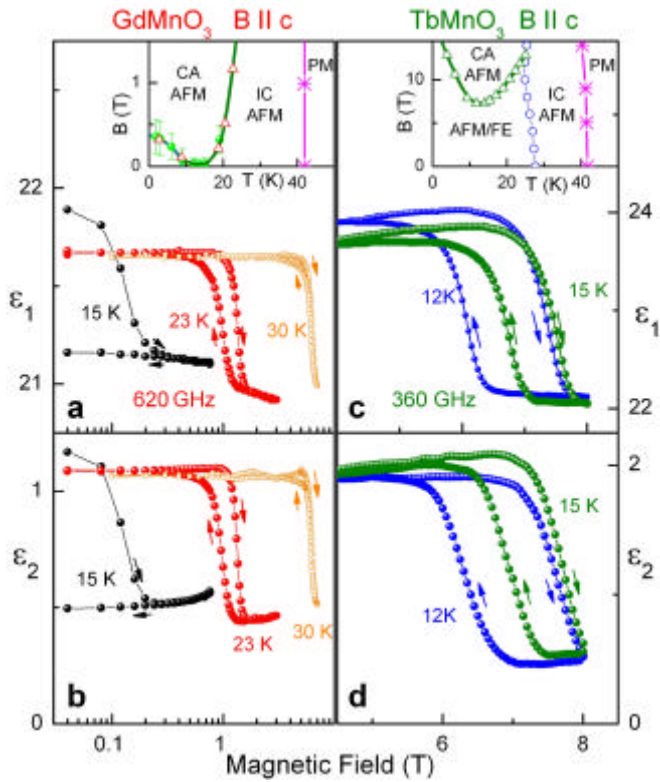

Fig. 1. Terahertz-magnetodielectric effect in $GdMnO_3$ and $TbMnO_3$. Real (**a,c**) and imaginary (**b,d**) parts of the a-axis dielectric constant in $GdMnO_3$ (**a,b**) and in $TbMnO_3$ (**c,d**) as a function of external magnetic field parallel to the c-axis and with the ac electric field parallel to a-axis. All data have been obtained in the zero-field-cooling starting from T = 25 K (IC-AFM phase). The insets show schematically the (B,T) phase diagrams of both compounds [12,11].

In order to clarify the physical mechanism of the magnetic-field induced changes in $GdMnO_3$ and $TbMnO_3$, we measured the frequency dependence of the dielectric properties both with and without external magnetic field. These results are presented in Fig. 2 and constitute the basic result of this work. The data obtained without magnetic field and with the electric ac-component e∥a exhibit a broad relaxation-like excitation with characteristic frequency $\nu_0 = 23 \pm 3$ cm$^{-1}$ in $GdMnO_3$ and $\nu_0 = 20 \pm 3$ cm$^{-1}$ in $TbMnO_3$ for all temperatures. The dielectric contribution of this excitation increases with decreasing temperature and saturates in the low-temperature magnetic phase. No significant changes are observed in $TbMnO_3$ (see the right inset of Fig. 1) when passing the magnetic phase boundary at 28 K although the damping of the mode decreases. We recall that this low-temperature phase in $TbMnO_3$ is ferroelectric with the polarization P∥c. Similar effects are observed in $GdMnO_3$. Here the IC-AFM phase remains stable with no induced ferroelectricity. From this observation is clear that, depending on symmetry, electromagnons can be observed in systems with inhomogeneous spin structure and ME coupling, independent of the existence of static ferroelectric polarization. It is unclear if the narrowing of the excitations results from the onset of Gd-ordering (T ≈ 8 K) in $GdMnO_3$ and of Tb-ordering (T ≈ 7 K) in $TbMnO_3$.

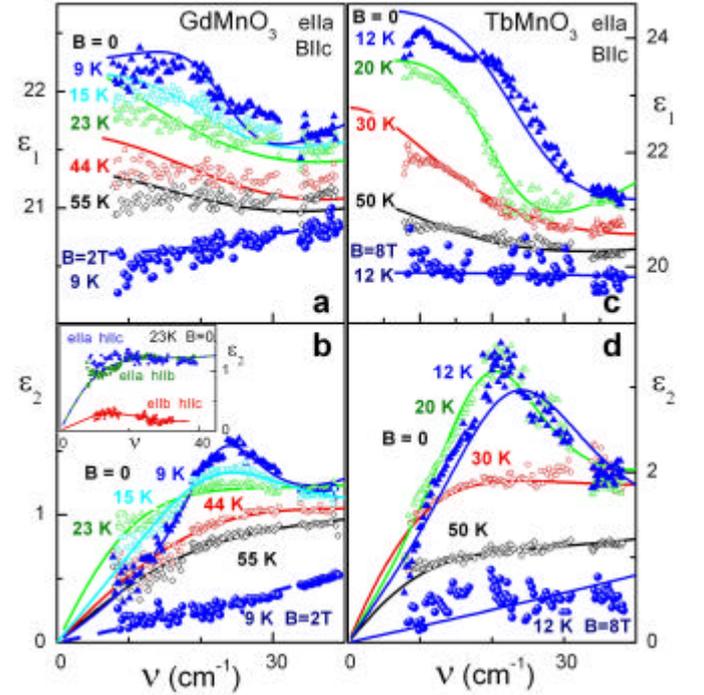

Fig. 2. Spectra of electromagnons in $GdMnO_3$ and $TbMnO_3$. Frequency dependence of the real (**a,c**) and imaginary (**b,d**) parts of the THz-dielectric function in $GdMnO_3$(**a,b**) and in $TbMnO_3$ (**c,d**) with e∥a and B∥c. Open symbols: experimental data in zero external magnetic field and in the incommensurate-antiferromagnetic phase. Solid lines represent model calculations adding an over-damped Lorentzian to the residual high-frequency contribution. Full spheres: the data in the canted antiferromagnetic state obtained applying B = 2 T ($GdMnO_3$) and B = 8 T ($TbMnO_3$) along the c-axis. The corresponding zero-field data are shown by full triangles. The inset demonstrates unique (e∥a, h - independent) excitation conditions for electromagnons.

The direct connection of the observed excitation with the magnetic subsystem can be immediately documented by applying a static magnetic field along the c-axis: The imaginary part of the dielectric constant (Fig.2**b,d**) is suppressed by more than a factor of two and the real part (Fig.2**a,b**) is reduced by about $\Delta\varepsilon_1 \sim 2$. This simultaneous suppression of $\varepsilon_1$ and $\varepsilon_2$ reflects the mutual connection of both quantities via the Kramers-Kronig relations.

The influence of an external magnetic field on the dielectric properties of GdMnO$_3$ and TbMnO$_3$ can be compared with the dramatic changes of infrared properties in colossal magnetoresistance manganites [15]. In the latter compounds the broad conductivity maximum at midinfrared frequencies (~0.5 eV) is transferred to the Drude-like relaxation at the metal-to-insulator transition (MIT). Especially in Ca-doped PrMnO$_3$ a MIT can be induced by magnetic fields [15], which leads to magnetoresistance of several orders of magnitude. In case of GdMnO$_3$ and TbMnO$_3$ the basic difference is that the transition in magnetic field takes place between two dielectric phases and only minimal changes in the low-frequency absorption are observed.

The unusual point concerning the observed mode in GdMnO$_3$ and TbMnO$_3$ is that it is excited by the *electric* ac-component of the radiation in spite of the clear connection of this mode to the modulated *magnetic* structure. To prove that this spin-wave excitation depend on the ac electric field and to determine the symmetry-allowed electromagnons we performed various experiments using the radiation with the electric and magnetic components polarized along all principal crystallographic axes. An impressive example of this behavior is documented in the inset of Fig. 2 which shows the suppression of the magnetic excitation when the electric ac-component is rotated from e∥a to e∥b leaving the magnetic field unchanged. On the contrary, this excitation remains unchanged if the ac magnetic field is rotated from h∥c to h∥b. This behaviour can be contrasted to the excitation of AFMR modes by magnetic ac-components only [16,17]. The results of the inset of Fig. 2 are fully compatible with the crystal symmetry and a modulated spin structure with propagation along the crystallographic b-direction. The sensitivity of the incommensurate mode to the ac electric field in GdMnO$_3$ and TbMnO$_3$ demonstrates strong coupling of magnetic and lattice degrees of freedom reflecting the close correlation of spin-structure and electric polarization.

Magnetodielectric (magnetoelectric) coupling plays a key role in exciting spin waves in a inhomogeneous modulated spin structure, since a homogeneous interaction is forbidden by symmetry arguments. The main contribution of this inhomogeneous magnetodielectric interaction [18] to the density of the free energy is determined by $F_{me} = -a_xP_x(A_x\partial A_y/\partial y - A_y\partial A_x/\partial y) - a_zP_z(A_z\partial A_y/\partial y - A_y\partial A_z/\partial y)$, where **P** is the electric polarization, **A** is the antiferromagnetic vector of the manganese spins, and $a_{x,z}$ are ME interaction constants. This form of $F_{me}$ is obtained using transformation properties of **A**, $\partial\mathbf{A}/\partial y$ and **P** according to eight irreducible representations (k=0) of the Pbnm crystallographic space-group. The gradient terms are nonzero for a modulated spin structure with a propagation vector $\mathbf{k} = (0, k_y, 0)$. Including the usual dielectric contribution $F_E = -\mathbf{PE} + \mathbf{P}^2/2\chi_E$, where $\chi_E$ is electric susceptibility and **E** is electric field, and minimizing by **P,** this ME interaction couples spin oscillations with the homogeneous ac electric field, contributes to the dielectric constant, and also can induce spontaneous electrical polarization in a modulated magnetic structure. An alternative form of the magnetodielectric coupling can be obtained by expansion of the free energy by **P** and magnetic order parameters $s_?(k)$ of the modulated structure with $\mathbf{k} = (0, k_y, 0)$ using their transformation with respect to irreducible representations $G_?$ of the propagation vector group $G_k$ [14]. Here spontaneous electrical polarization in a modulated magnetic structure with a strong dependence on symmetry can also be induced (e.g., $P_z \sim \langle A_z\partial A_y/\partial y - A_y\partial A_z/\partial y\rangle_{average}$). Interestingly, the spin oscillations excited by the electric field correspond to non-zero wave vectors **k** of spin waves, and the resonance frequencies of electrically and magnetically excited modes are well separated in energy.

In conclusion, we have shown that in magneto-dielectric materials fundamentally new hybrid spin-lattice excitations exist which can be excited by ac-electric fields. The appearance/disappearance of this new type of excitations comes along with significant changes in the index of refraction. Hence, electromagnons are not only interesting for basic research but also for the design of magnetooptic devices.

**Methods**

Single crystals of GdMnO$_3$ and TbMnO$_3$ have been prepared by floating-zone method with radiation heating. The samples have been characterized using X-ray, electrical, magnetic and thermodynamic measurements. The dielectric properties agree well with published results [2,12].

For T-ray experiments (0.1-1.2 THz) various plane-parallel plates were cut from the original boules each oriented perpendicular to one of the principal crystallographic axes. Typical size of the samples was 5×5×1 mm$^3$. Different geometries were necessary to distinguish between electrical (i.e. excited by the ac-component of the electrical field of the electromagnetic radiation) and magnetic modes and allowed proving that the observed mode of the incommensurate AFM structure indeed is excited by the electric component of the radiation.

The dynamic experiments in the frequency range 0.1< ν < 1.2 THz have been performed in transmittance experiments using a Mach-Zehnder interferometer [19]. This arrangement allows to investigate the transmittance and phase-shift of the plane-parallel samples as function of frequency, temperature, and external magnetic field. Dynamic dielectric properties $\varepsilon^*(\omega,T,B) = \varepsilon_1 + i\varepsilon_2$ were calculated from these quantities using the Fresnel optical equation for the complex transmission coefficient of the plane-parallel plate [20] without additional assumptions. Temperature-dependent and magnetic-field experiments have been carried out in a split-coil magnet providing magnetic fields up to 7 T and temperatures from 1.8 K ≤ T ≤ 300 K.

**Acknowledgements** The stimulating discussion with J. Hemberger and P. Lunkenheimer and M. Kenzelmann is gratefully acknowledged. We thank T. Kimura for sharing with us the data and samples of TbMnO$_3$ and A. Pimenova for performing magnetization experiments. This work was supported by BMBF (13N6917/0 - EKM), by DFG (SFB484 - Augsburg) and by RFBR (03-02-16759, 06-02-17514).

**Correspondence** and requests for materials should be addressed to A. P. (andrei.pimenov@physik.uni-augsbug.de)